\documentclass[aps,prl,twocolumn,preprintnumbers,showpacs,nofootinbib]{revtex4}

\usepackage[english]{babel}
\usepackage{amsmath}
\usepackage{amssymb}
\usepackage{epsfig}
\usepackage{graphics,psfrag,rotating}
\usepackage{graphicx}% Include figure files
\usepackage{dcolumn}% Align table columns on decimal point
\usepackage{bm}% bold math
\begin{document}
\newcommand{\Od}{{\cal O}}
\newcommand{\lsim}   {\mathrel{\mathop{\kern 0pt \rlap
  {\raise.2ex\hbox{$<$}}}
  \lower.9ex\hbox{\kern-.190em $\sim$}}}
\newcommand{\gsim}   {\mathrel{\mathop{\kern 0pt \rlap
  {\raise.2ex\hbox{$>$}}}
  \lower.9ex\hbox{\kern-.190em $\sim$}}}

%\preprint{APS/123-QED}

\title{On the evolution of density perturbations in $f(R)$
 theories of gravity}%
\author{A. de la Cruz-Dombriz\footnote{E-mail: dombriz@fis.ucm.es}, 
A. Dobado\footnote{E-mail:dobado@fis.ucm.es} and
A. L. Maroto\footnote{E-mail: maroto@fis.ucm.es}}

\affiliation{Departamento de  F\'{\i}sica Te\'orica I, Universidad 
Complutense de Madrid, E-28040 Madrid, Spain}%
%\affiliation{...Minessota}
\date{\today}% It is always \today, today,
%\date{November $15^{th}$ 2007}

\begin{abstract}
In the context of $f(R)$ theories of gravity,
we study the evolution of scalar 
cosmological perturbations in the metric formalism. 
Using a completely general procedure, we find the exact fourth-order 
differential equation for the matter density perturbations in 
the longitudinal gauge. 
In the case of sub-Hubble modes, the expression reduces  to a second-order 
equation which is compared with the standard (quasi-static)
equation used in the literature.  We show that for general
$f(R)$ functions the quasi-static approximation is not justified. However, 
for those  functions adequately describing the present phase of 
accelerated expansion and  satisfying local gravity tests, 
it provides a correct description for the evolution of perturbations.  
\end{abstract}

\maketitle

\section{Introduction}
The present phase of accelerated expansion of the universe \cite{SN} poses one 
of the most important problems of modern cosmology. It is well known that
ordinary  Einstein's equations in either a matter or radiation dominated universe
give rise to decelerated periods of expansion. In order to have acceleration, 
the total energy-momentum tensor appearing on the right hand side of
the equations should be dominated at late times by a hypothetical negative 
pressure fluid  usually called dark energy (see \cite{review} 
and references therein).

However, there are other possibilities to generate a period of acceleration 
in
which  no new sources are included on the r.h.s. of the equations, but instead 
Einstein's gravity itself is modified \cite{otras}. 
In one of such possibilities,  new functions
of the curvature scalar ($f(R)$ terms)  are included  
in the gravitational action, which amounts to modifiying the l.h.s of the 
equations of motion. Although such theories are able to describe
the accelerated expansion on cosmological scales correctly, they typically 
give rise to 
strong effects on smaller scales. In any case  viable models 
can be constructed to be compatible with local gravity tests and other
cosmological constraints \cite{varia}.

The important question that  arises  is therefore  
how to discriminate dark energy models from 
modified gravities using present or future observations.
It is known that by choosing particular $f(R)$ functions, 
one can mimic any background evolution 
(expansion history), and in particular that of  $\Lambda$CDM.
Accordingly, the exclusive use of observations such as 
high-redshift Hubble diagrams from SNIa \cite{SN}, baryon acoustic
oscillations \cite{BAO} or CMB shift factor \cite{WMAP}, 
based on different distance measurements
which are sensitive only to the expansion history, cannot settle the question
of the nature of dark energy \cite{Linder2005}.

However, there exists a different type of observations 
which are sensitive, not 
only to the expansion history, but also to the evolution of matter 
density perturbations. The fact that the evolution of perturbations 
depends on the specific 
gravity model, i.e. it differs in general from
that of Einstein's gravity even though the background evolution is the same, 
means
that this kind of observations will help distinguishing between different  
models for acceleration.

In this work we study the problem of determining the exact equation 
for the evolution of matter density perturbations for arbitrary $f(R)$
theories. Such problem had been previously considered in the literature 
(\cite{perturbations,Silvestri_09_2007,Qstatic_Zhang,Qstatic_Bean,Amendola,
Starobinsky2})  
and approximated equations have been widely used. They are typically based
on the so called quasi-static approximation in which  all the  time derivative
terms for the gravitational potentials are discarded, and only those
including density perturbations are kept \cite{Starobinsky}. From our exact 
result, we will be able to determine under which conditions such an 
approximation 
can be justified. 

The paper is organized as follows: in Section 2, we
briefly review the perturbations equations for the standard $\Lambda$CDM
model. In Section 3 we obtain the perturbed equations for general 
$f(R)$ theories.
In Section 4 we describe the procedure to obtain the general equation
for the density perturbation.  In Section 5 
we summarize the main viability condition for $f(R)$ theories. Section 6 
is devoted to the study of the 
validity of the quasi-static approximation. In Section 7 we apply our
results to some particular models and finally in Section 8 we include the
main conclusions. In Appendices I and II 
we have also included complete expressions for the 
relevant coefficients of the perturbation equation.

\section{Density perturbations in $\Lambda$CDM}
Let us start by considering the simplest model for dark energy described 
by a cosmological constant $\Lambda$. The corresponding Einstein's
equations read:
\begin{eqnarray}
G^\mu_{\;\;\nu}=-8\pi \text{G} T^{\mu}_{\;\;\nu} - 
\Lambda \delta^{\mu}_{\;\;\nu}
\end{eqnarray}
where $G^\mu_{\;\;\nu}$ is the Einstein's tensor and $T^{\mu}_{\;\;\nu}$
is the energy-momentum tensor for matter.

In the metric formalism for the $\Lambda \text{CDM}$ model 
it is possible to obtain a second order differential equation for 
the growth of matter density perturbation 
$\delta\,\equiv\,\delta\rho/\rho_{0}$. Let us consider the scalar perturbations
of a flat FRW metric in the longitudinal gauge and in conformal time: 
\begin{equation}
ds^2\,=\,a^2(\eta)[(1+2\Phi)d\eta^2-(1-2\Psi)(dr^2+r^2d\Omega_{2}^2)] 
\label{perturbed_metric}
\end{equation}
where $\Phi\equiv\Phi(\eta,\overrightarrow{x})$ and 
$\Psi\equiv\Psi(\eta,\overrightarrow{x})$ are the scalar perturbations. 
From this metric, we obtain the first-order perturbed Einstein's equation:
\begin{eqnarray}
\delta G^{\mu}_{\;\;\nu}=-8\pi \text{G}\delta T^{\mu}_{\;\;\nu} 
\end{eqnarray}
where the perturbed energy-momentum tensor reads:
\begin{eqnarray}
\delta T^{0}_{\;\;0}\,&=&\,\delta\rho = \rho_{0} \delta \,\, , \,\, 
\delta T^{i}_{\;\;j}\,=\, -\delta P \delta^{i}_{\;\;j} = 
-c_{s}^2 \delta^{i}_{\;\;j}\rho_{0}\delta \nonumber\\ 
\delta T^{0}_{\;\;i}\,&=&\,-\delta T^{i}_{\;\;0}\,=
-\,(1 + c_{s}^2)\rho_{0}\partial_{i}v  
\label{perturbed_stress_energy_tensor}
\end{eqnarray}
with $\rho_0$ the unperturbed energy density and $v$ the potential 
for velocity perturbations. We assume that the perturbed and unperturbed 
matter have the same equation of state, i.e. 
$\delta P/\delta\rho\equiv c_{s}^2 
\equiv P_{0}/\rho_{0}$,   
where $c_s=0$ for matter perturbations.  
The resulting differential equation for $\delta$ in Fourier space 
is written as:
\begin{eqnarray}
&&\delta''+\mathcal{H}\frac{k^4 -6\tilde{\rho}k^2 -18\tilde{\rho}^2}
{k^4 - \tilde{\rho}(3 k^2 +9\mathcal{H}^2)}\,\delta'-\nonumber\\
&&-\tilde{\rho}\frac{k^4 + 9\tilde{\rho}(2\tilde{\rho}-3\mathcal{H}^2) 
-k^2 (9\tilde{\rho}-3\mathcal{H}^2)}{k^4 - \tilde{\rho}(3 k^2 
+9\mathcal{H}^2)}\,\delta\,=\,0
\label{delta_0_general_LambdaCDM}
\end{eqnarray}
where  $\tilde{\rho} \equiv 4\pi \text{G} \rho_{0} a^2 
\,=\, -\mathcal{H}'+\mathcal{H}^2$ and $\mathcal{H}\equiv a'/a$ with  prime 
denoting derivative with respect to time $\eta$. We point out that  
it is not necessary to explicitly calculate potentials $\Phi$ and $\Psi$ 
to obtain equation \eqref{delta_0_general_LambdaCDM}, but algebraic 
manipulations in the field  
equations are enough to get this result. In the extreme sub-Hubble limit, 
i.e. $k\eta\gg \,1$ or equivalently $k\gg\,\mathcal{H}$, 
\eqref{delta_0_general_LambdaCDM} is reduced to the well-known expression:
\begin{equation}
\delta''+\mathcal{H}\delta' - 4\pi\text{G} \rho_0 a^2\delta\,=\,0
\label{delta_0_SubHubble}
\end{equation}

In this regime and at early times, the matter energy density dominates over
 the cosmological constant and it is easy to show 
that $\delta$ solutions for \eqref{delta_0_SubHubble} grow as
$a(\eta)$. At late times 
(near today) the cosmological constant contribution is not negligible and 
power-law solutions for \eqref{delta_0_SubHubble}  no longer exist. 
It is necessary in this case to 
assume an ansatz for $\delta$. One which works very well is the 
one proposed in \cite{Linder2005} and \cite{Linder2006} 
\begin{equation}
\frac{\delta(a)}{a}\,= \, e^{\int_{a_{i}}^{a}[\Omega_{m}(a)^{\gamma}-1]\text{d}\,\text{ln} a} 
\label{modelo_delta_Linder}
\end{equation}
This expression fits with high precission the numerical solution for 
$\delta$  with a constant parameter $\gamma\,=\,6/11$.

\section{Perturbations in $f(R)$ theories}
Let us consider the modified gravitational  action:
\begin{eqnarray}
S=\frac{1}{16\pi \text{G}}\int d^4x \sqrt{-g}\left(R+f(R)\right)
\label{action}
\end{eqnarray}
where $R$ is the scalar curvature\footnote{The Riemann tensor definition 
is $R^\mu_{\;\;\nu\alpha\beta}=\partial_\beta\Gamma^{\mu}_{\nu\alpha}
-\partial_\alpha\Gamma^{\mu}_{\nu\beta}
+\Gamma^{\mu}_{\sigma\beta}\Gamma^{\sigma}_{\nu\alpha}
-\Gamma^{\mu}_{\sigma\alpha}\Gamma^{\sigma}_{\nu\beta}$
which has opposite sign to the one proposed in \cite{Giovannini}.}.
The corresponding equations of motion read:
\begin{eqnarray}
G_{\mu\nu}&-&\frac{1}{2}g_{\mu\nu}f(R)+R_{\mu\nu}f_{R}(R)-g_{\mu\nu}\Box
f_{R}(R)+f_{R}(R)_{;\mu\nu}\nonumber \\
&=&-\,8\pi \text{G} T_{\mu\nu}
             %%% ¡ojo! %%%%
%% He cambiado de signo los dos últimos sumandos del miembro izquierdo. %%%%
%%% EN EL MIEMBRO DERECHO SE PONE ``MENOS`` 8\pi... %%%%%%%%%%%%%%%%%%%%%%%%
%%%%%%%%%%%%%%%%%%%%%%%%%%%%%%%%%%%%%%%%%%%%%%%%%%%%%%%%%%%%%%%%%%%%%%%%%%%%
\end{eqnarray}
where $f_{R}(R)=df(R)/dR$.
For the background flat Robertson-Walker metric they read:
\begin{equation}
\frac{3\mathcal{H}'}{a^2}(1+f_{R})-\frac{1}{2}(R_{0}+f_{0})
-\frac{3\mathcal{H}}{a^2}f_{R}' \,=\,-8\pi \text{G}\rho_{0}
\label{00_bg_eqn}
\end{equation}
and
\begin{eqnarray}
\frac{1}{a^2}(\mathcal{H}'&+&2\mathcal{H}^2)(1+f_{R})
-\frac{1}{2}(R_{0}+f_{0})-\frac{1}{a^2}(\mathcal{H} f_{R}'+f_{R}'') \,
\nonumber \\
&=&\,8\pi \text{G} c_{s}^2 \rho_{0}
\label{ii_bg_eqn}
\end{eqnarray}
where $R_{0}$ denotes the  scalar curvature corresponding to the  
unperturbed metric, 
$f_{0}\equiv f(R_0)$, $f_{R} \equiv \text{d}f(R_0)/\text{d}R_{0}$ and  
prime means derivative with respect to time $\eta$. 
A very useful equation to use in the following calculations 
is the \eqref{ii_bg_eqn} $-$ \eqref{00_bg_eqn} combination
\begin{eqnarray}
2(1+f_{R})(-\mathcal{H}'+\mathcal{H}^2)+2\mathcal{H}f_{R}'
-f_{R}''\,=\,8\pi \text{G}\rho_{0}(1+c_{s}^2) a^2\nonumber \\ 
\label{bg_density_plus_pression}
\end{eqnarray}
 Finally we have the conservation equation:
\begin{equation}
\begin{array}{l}
\rho_{0}'+3(1+c_{s}^2)\mathcal{H}\rho_0\,=\,0 
\end{array}
\label{bg_density_conservation}
\end{equation}

Using the perturbed metric (\ref{perturbed_metric}) and
the perturbed energy-momentum tensor (\ref{perturbed_stress_energy_tensor}), 
the first order perturbed equations, 
assuming  that the  background equations  hold, may be written as:
\begin{eqnarray}
(1+f_{R})\delta G^{\mu}_{\nu}+(R_0\,^{\mu}_{\nu} +\nabla^{\mu}\nabla_{\nu}
-\delta^{\mu}_{\nu} \square)f_{RR}\delta R +
\nonumber\\  
\,[(\delta g^{\mu\alpha}) \nabla_{\nu}\nabla_{\alpha} - \delta^{\mu}_{\nu} 
(\delta g^{\alpha\beta})\nabla_{\alpha}\nabla_{\beta}]f_{R}-
\nonumber\\  
\,[g_{0}^{\alpha\mu}(\delta\Gamma^{\gamma}_{\alpha\nu})
-\delta^{\mu}_{\nu} g_{0}^{\alpha\beta}
(\delta\Gamma^{\gamma}_{\beta\alpha})] \partial_{\gamma}f_{R}\,=
\,-8\pi G \delta T^{\mu}_{\nu}
\label{tensorial_bg_eqns}
\end{eqnarray}
where $f_{RR}\,=\,\text{d}^{2}f(R_{0})/\text{d}R_{0}^2$, $\square\,\equiv\,\nabla_{\alpha}\nabla^{\alpha}$ and $\nabla$ 
is the usual covariant derivative with respect to the unperturbed FRW metric 
(see \cite{Giovannini} for perturbed metric, connection symbols and other 
useful perturbed quantities). Notice that unlike the 
ordinary Einstein-Hilbert case, with second order equations, this is a set of
 fourth-order differential
equations.
By computing the covariant 
derivative with respect to the perturbed metric $\tilde{\nabla}$ 
of the perturbed energy-momentum tensor $\tilde{T}^{\mu}_{\nu}$, 
we find the conservation equations:
\begin{equation}
\tilde{\nabla}_{\mu}\tilde{T}^{\mu}_{\nu}\,=\,0
\label{cons}
\end{equation} 
which do not depend on $f(R)$.
 
For the linearized Einstein's equations, the components  
$(00)$, $(ii)$, $(0i)\equiv (i0)$ and $(ij)$, where $i,j\,=\,1,2,3$, 
$i\neq j$, in  Fourier space, read respectively:
\begin{eqnarray}
(1+f_{R})&[&-k^2(\Phi+\Psi)-3\mathcal{H}(\Phi'+\Psi')+(3\mathcal{H}'
-6\mathcal{H}^2)\Phi-
\nonumber\\
3\mathcal{H}'\Psi&]&+f'_{R}
(-9\mathcal{H}\Phi+3\mathcal{H}\Psi-3\Psi')\,=\,2\tilde{\rho}\delta
\label{00_pert}
\end{eqnarray}
\begin{eqnarray}
(1+f_{R})&[&\Phi''+\Psi''+3\mathcal{H}(\Phi'+\Psi')
+3\mathcal{H}'\Phi+(\mathcal{H}'+
\nonumber\\
2\mathcal{H}^2)\Psi&]&
+f'_{R}(3\mathcal{H}\Phi-\mathcal{H}\Psi+3\Phi') + f''_{R}(3\Phi-\Psi)\,=\,
\nonumber\\
2c_{s}^{2} \tilde{\rho}\delta
\label{ii_pert}
\end{eqnarray}
\begin{eqnarray}
(1+f_{R})[\Phi'+\Psi'+\mathcal{H}(\Phi+\Psi)]+f'_{R}(2\Phi-\Psi)\,=\,
\nonumber\\
- 2\tilde{\rho}(1+c_{S}^2) v
\label{0i_pert}
\end{eqnarray}
\begin{eqnarray}
\Phi-\Psi\,=\,-\frac{f_{RR}}{1+f_{R}}\delta R
\label{ij_pert}
\end{eqnarray}
where $\delta R$ is given by:
\begin{eqnarray}
\delta R \,=\, &-&\frac{2}{a^2}[3\Psi''+6(\mathcal{H}'+\mathcal{H}^2)\Phi
+3\mathcal{H}(\Phi'+3\Psi')-
\nonumber\\
&&k^2(\Phi-2\Psi)]
\label{deltaRdefinition}
\end{eqnarray}

Finally, from the energy-momentum tensor conservation (\ref{cons}), we get
to first order: 
\begin{eqnarray}
3\Psi'(1+c_{s}^2)-\delta'+k^2(1+c_{s}^2)v\,=\,0
\label{Nabla0}
\end{eqnarray}
and 
\begin{eqnarray}
\Phi+\frac{c_{s}^2}{1+c_{s}^2}\delta+v'+\mathcal{H}v(1-3c_{s}^2)\,=\,0
\label{Nablaj}
\end{eqnarray}
for the temporal and spatial components respectively.

In a dust matter dominated universe, i.e. $c_{s}^2\,=\,0$, 
\eqref{Nabla0} and \eqref{Nablaj} can be combined to give
\begin{eqnarray}
\delta''+\mathcal{H}\delta'+k^2\Phi-3\Psi''-3\mathcal{H}\Psi'\,=\,0
\label{Nabla}
\end{eqnarray}
which will be very useful in future calculations.

\section{Evolution of density perturbations}

Our pourpose is to derive a fourth order differential equation 
for matter density perturbation $\delta$ alone. This can be performed 
by means of the following process:

Let us consider equations \eqref{00_pert} and \eqref{0i_pert} for a 
matter dominated universe i.e. $c_{s}^2=0$, and combine them 
to express the potentials $\Phi$ and $\Psi$ in terms of 
$\{ \Phi', \Psi', \delta , \delta' \}$ by means of 
algebraic manipulations. The resulting expressions are the following
\begin{eqnarray}
\Phi\,&=&\,\frac{1}{\mathcal{D}(\mathcal{H}, k)}\Big\{[3(1 +f_{R})\mathcal{H}(\Psi'+\Phi') \nonumber\\
&+& f_{R}'\Psi' +2\tilde{\rho} \delta](1+f_{R})(\mathcal{H}-f_{R}')+ [(1+f_{R})(\Phi'+\Psi') \nonumber\\
&+&\frac{2\tilde{\rho}}{k^2}(\delta'-3\Psi')][(1+f_{R})(-k^2-3\mathcal{H}') + 3f_{R}'\mathcal{H}]\Big\}
\label{Phi_desp}
\end{eqnarray}
and
\begin{eqnarray}
\Psi\,&=&\,\frac{1}{\mathcal{D}(\mathcal{H},k)}\Big\{[-3(1+f_{R})\mathcal{H}(\Psi'+\Phi')-3f_{R}'\Psi'\nonumber\\
&-&2\tilde{\rho}\delta][(1+f_{R})\mathcal{H}+2f_{R}']-[(1+f_{R})(\Phi'+\Psi')\nonumber\\
&+&\frac{2\tilde{\rho}}{k^2}(\delta'-3\Psi')][(1+f_{R})(-k^2+3\mathcal{H}'-6\mathcal{H}^2) -9\mathcal{H}f_{R}']\Big\}\nonumber\\
\label{Psi_desp}
\end{eqnarray}
where 
\begin{eqnarray}
\mathcal{D}(\mathcal{H},k)\,&\equiv&\,-6(1+f_{R})^2\mathcal{H}^{3}+3\mathcal{H}[f_{R}'^{2}+2(1+f_{R})^2\mathcal{H}']+\nonumber\\
&&3(1+f_{R})f_{R}'(-2\mathcal{H}^{2}+k^2+\mathcal{H}')
\label{Phi&Psi_denominator}
\end{eqnarray}
The second step will be to derive equations 
\eqref{Phi_desp} and \eqref{Psi_desp} with respect to 
$\eta$ and obtain $\Phi'$ and $\Psi'$ algebraically in 
terms of $\{\Phi'',\Psi'';\delta,\delta',\delta'' \}$.  
These last results can be substituted in 
equations \eqref{00_pert} and \eqref{0i_pert} to 
obtain potentials $\Phi$ and $\Psi$ just in terms 
of  $\{\Phi'',\Psi'',\delta,\delta',\delta''\}$. So at this stage 
we are able to express, but we do not do here explicitly, the following
\begin{eqnarray}
\Phi \,&=&\,\Phi(\Phi'',\Psi''; \delta,\delta',\delta'') \nonumber\\
\Psi\,&=&\,\Psi(\Phi'',\Psi''; \delta,\delta',\delta'') \nonumber\\
\Phi'\,&=&\,\Phi'(\Phi'',\Psi''; \delta,\delta',\delta'') \nonumber\\ \Psi'\,&=&\,\Psi'(\Phi'',\Psi'';\delta,\delta',\delta'')
\label{Todos}
\end{eqnarray}
where we mean that the functions on the l.h.s. are algebraically 
dependent on the functions inside the parenthesis on the r.h.s.

The natural reasoning at this point would be to try to obtain the 
potentials second derivatives $\{\Phi'',\Psi''\}$ in terms 
of $\{\delta,\delta',\delta''\}$ by an algebraic process. The 
chosen equations to do so will be \eqref{Nabla} 
and \eqref{ij_pert} first derivative with respect to $\eta$. 
In \eqref{Nabla}  it is necessary to substitute $\Phi$ 
and $\Psi'$ by the expressions obtained in \eqref{Todos} 
whereas \eqref{ij_pert} first derivative may be sketched as follows
\begin{equation}
\Phi'-\Psi'= - \frac{f_{RR}}{1+f_{R}}\delta R' 
+ \Big[\frac{f_{RR}f_{R}'-f_{RR}'(1+f_{R})}{(1+f_{R})^2}\Big]\delta R 
\label{ij_pert_derivative}
\end{equation}
Before deriving, we are going to substitute 
$\Psi''$ that appears on \eqref{ij_pert} by 
lower derivatives potentials $\{\Phi,\Psi,\Phi', \Psi'\}$, $\delta$ 
and its derivatives. To do so we consider 
\eqref{00_pert} and \eqref{0i_pert} first derivatives with 
respect to $\eta$ where the quantity $v$ has been previously 
substituted by its expression in \eqref{Nabla0}. Following 
this process we may express $\Psi''$ as follows
\begin{eqnarray}
\Psi''\,=\,\Psi''(\Phi, \Psi, \Phi', \Psi'; \delta, \delta', \delta'')
\label{Psi2_function}
\end{eqnarray}
and now substituting in \eqref{ij_pert} we can derive that 
equation with respect to $\eta$. Solving a two 
algebraic equations system with equations 
\eqref{Nabla} and \eqref{ij_pert_derivative} and 
introducing \eqref{Todos} we are able to express 
$\{\Phi'',\Psi''\}$ in terms of $\{\delta,\delta',\delta'',\delta'''\}$.
\begin{eqnarray}
\Phi''\,=\,\Phi''(\delta,\delta',\delta'',\delta''') \,\,;\,\,  \Psi''\,=\,\Psi''(\delta,\delta',\delta'',\delta''')
\label{Phi2&Psi2_fuentes}
\end{eqnarray}

We substitute the results obtained in 
\eqref{Phi2&Psi2_fuentes} straightforwardly in \eqref{Todos} 
in order to express $\{\Phi$, $\Psi$, $\Phi ',\Psi'\}$ 
in terms of $\{\delta ,\delta',\delta'',\delta'''\}$. With the 
two potentials and its first derivatives as algebraic functions 
of $\{\delta, \delta', \delta'',\delta'''\}$, we performe the last step: 
We consider $\Phi(\delta,\delta,\delta'',\delta''')$ and derive it with 
respect to $\eta$. The result should be equal 
to $\Phi'(\delta,\delta,\delta'',\delta''')$ so we only need to express 
together these two results obtaining a fourth order differential 
equation for $\delta$. Note that  this procedure 
is completely general to first order for 
scalar perturbations in the metric formalism for $f(R)$ gravities.

Once this fourth order differential equation has been 
solved we may go backwards and by using the results for  
$\delta$  we obtain  $\{\Phi'', \Psi''\}$ 
from \eqref{Phi2&Psi2_fuentes} as functions of time. 
Analogously from \eqref{Todos} the behavior of the 
 potentials $\{\Phi,\Psi\}$ and their first derivatives could be 
determined. 

The resulting equation for $\delta$  can be written as follows:
\begin{eqnarray}
\beta_{4,f}\delta^{iv}+\beta_{3,f}\delta'''
&+&(\alpha_{2,\text{EH}}+\beta_{2,f})\delta''
+(\alpha_{1,\text{EH}}+\beta_{1,f})\delta'+ \nonumber\\
(\alpha_{0,\text{EH}}+\beta_{0,f})\delta \,&=&\,0
\label{delta_equation_separated}
\end{eqnarray}
where the coefficients $\beta_{i,f}$  $(i\,=\,1,...,4)$ 
involve terms with $f_{R}'$ and $f_{R}''$, i.e. 
terms disappearing if we take $f_{R}$ constant. 
Equivalently,  $\alpha_{i,\text{EH}}$ $(i\,=\,0,1,2)$ contain 
terms coming from the linear part of $f_{0}$ in $R_0$. 

It is very useful to define the parameter 
$\epsilon \equiv \mathcal{H}/k$ since it will  allow us  to 
perform a perturbative expansion of the previous coefficients 
$\alpha$'s and $\beta$'s in the sub-Hubble limit. 
Other dimensionless parameters which will be used are 
the following:  
$\kappa_{i} \equiv \mathcal{H}^{'^{(i)}}/\mathcal{H}^{i+1}$ ($i=1,2,3$) 
and $f_{i}\equiv f_{R}^{'^{(j)}}/(\mathcal{H}^{j} f_{R})$ ($j=1,2$). 

Expressing the $\alpha$'s and $\beta$'s coefficients with 
those dimensionless quantities we may write
\begin{eqnarray}
\alpha_{i,\text{EH}}\,&=&\,\sum_{j=1}^{3}\alpha^{(j)}_{i,\text{EH}}\,\,\,\,\, i=0,1,2 \nonumber\\
\beta_{i,f}\,&=&\,\sum_{j=1}^{7}\beta^{(j)}_{i,f} \,\,\,\,\, i=3,4
\nonumber\\
\beta_{i,f}\,&=&\,\sum_{j=1}^{8}\beta^{(j)}_{i,f} \,\,\,\,\, i=0,1,2
\end{eqnarray}
where two consecutives terms in each serie differ in $\epsilon^2$ factor. 
The expressions for the coefficients are too long to be written
explicitly. Instead, in the following sections we will show different 
approximated
formulae useful in certain limits.

\section{Viable $f(R)$ theories}

Results obtained so far are valid for any $f(R)$ theory. However, as mentioned in the introduction, this kind of models are severely constrained
in order to provide consistent theories of gravity. In this section 
we review the main conditions \cite{Silvestri_09_2007}:

1. $f_{RR}>0$ for high curvatures \cite{Hu&Sawicki_May_2007}. 
This is the requirement for a 
classically stable high-curvature regime and the existence of a matter 
dominated phase in the cosmological evolution.

2. $1+f_{R}>0$ for all $R_0$. This condition ensures the effective Newton's
constant to be positive at all times and the graviton energy to be positive.

3. $f_{R}<0$ ensures ordinary General Relativity behaviour is recovered at
early times. Together with the condition $f_{RR}>0$, it implies that 
$f_{R}$ should be negative and monotonically growing function of $R_0$
in the range $-1<f_{R}<0$.

4. $\vert f_{R}\vert\ll 1$  at recent epochs. This is imposed 
by local gravity tests \cite{Hu&Sawicki_May_2007}, although it is still not clear what is the actual 
limit on this parameter. This condition also implies that the cosmological 
evolution at late times resembles 
that of $\Lambda$CDM. In any case, this constraint is not required if we are
only interested in building models for cosmic acceleration. 

\section{Evolution of sub-Hubble modes and the quasi-static approximation} 
We are interested in the possible effects on the growth of 
density perturbations once they enter the Hubble radius in the matter dominated era. 
In the sub-Hubble limit $\epsilon\ll 1$, it can be seen that the 
$\beta_{4,f}$ and $\beta_{3,f}$ coefficients are supressed by $\epsilon^2$
with respect to  $\beta_{2,f}$, $\beta_{1,f}$ and $\beta_{0,f}$, i.e., 
in this limit the equation for perturbations reduces to the 
following  second order expression:
 \begin{widetext}
\begin{eqnarray}
\delta''+\mathcal{H}\delta'+\frac{(1+f_{R})^{5} \mathcal{H}^{2} 
(-1+\kappa_1)(2\kappa_1-\kappa_2)-\frac{16}{a^8}
 f_{RR}^{4}(\kappa_2-2)k^{8}8\pi \text{G} \rho_{0}a^2}
{(1+f_{R})^{5}(-1+\kappa_1)+\frac{24}{a^8}f_{RR}^{4}(1+f_{R})
(\kappa_2-2)k^{8}}\delta \,=\,0
\label{eqn_ours}
\end{eqnarray}
\end{widetext} 
where we have taken only the leading terms in the $\epsilon$ expansion for
the $\alpha$ and $\beta$ 
coefficients.

This expression can be compared with that  usually  
considered in literature, obtained after performing strong 
simplifications in the perturbed equations - \eqref{00_pert}, 
\eqref{ii_pert}, \eqref{0i_pert}, \eqref{ij_pert}, 
\eqref{Nabla0} and  \eqref{Nablaj} - by 
neglecting time derivatives of $\Phi$ and $\Psi$ potentials, 
(see  \cite{Starobinsky}).
Thus in \cite{Qstatic_Zhang} and \cite{Qstatic_Tsujikawa} they obtain:
\begin{eqnarray}
\delta^{''}+\mathcal{H}\delta^{'}-\frac{1+4\frac{k^2}{a^2}
\frac{f_{RR}}{1+f_{R}}}{1+3\frac{k^2}{a^2}\frac{f_{RR}}{1+f_{R}}}
\frac{\tilde{\rho}\delta}{1+f_{R}} \,=\,0
\label{Qstatic_strong_equation}
\end{eqnarray}
This approximation has been considered as too aggressive in \cite{Qstatic_Bean} since neglecting time derivatives can remove important information about the evolution. 

Note also that there exists a difference in a 
power $k^8$ between those terms coming from the $f$ part and those coming 
from the $EH$ part in (\ref{eqn_ours}). 
This result differs from that in the quasi-static 
approximation where difference is in a power $k^2$ according to \eqref{Qstatic_strong_equation}.

\begin{figure}[h]
\begin{center}
%\epsfxsize=10cm   %width of figure - will enlarge/reduce the figures
%\epsfbox{fig3.eps}
%\figurebox{2cm}{3cm}{} %to have a box alone
\resizebox{8.8cm}{6.4cm} 
%\resizebox{8.5cm}{!}
{\includegraphics{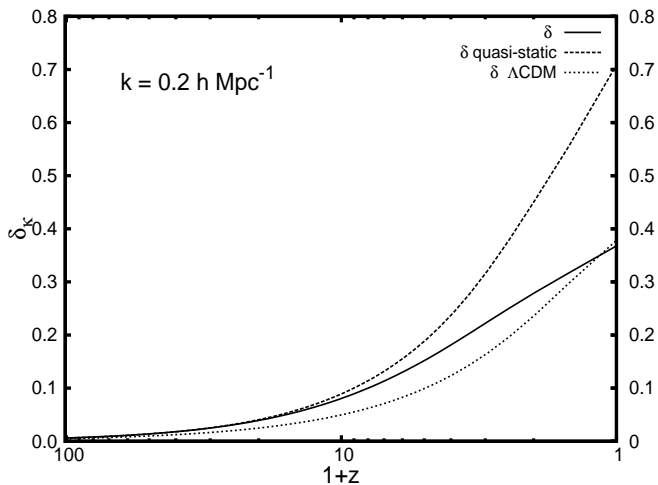}}
\caption {\footnotesize $\delta_{k}$ with $k = 0.2 \text{h}\text{Mpc}^{-1}$ 
for $f_{test}(R)$ model and $\Lambda\text{CDM}$. Both, standard 
quasi-static evolution and equation \eqref{eqn_ours} have 
been plotted in the redshift range from 100 to 0.}
\label{Figure_Models_delta_Falso}
\end{center}
\end{figure}

In order to compare the evolution for both equations, 
we have considered a specific function 
$f_{test}(R)=-4R^{0.63}$, where 
$H_0^2$ units have been used, which gives rise to a matter era 
followed by  a late time
accelerated phase with the correct deceleration 
parameter today. Initial conditions in the matter era 
were given at redshift $z =485$ 
where the $EH$ part was  dominant. Results, 
for $k=600 H_{0}$ are presented in figure (1).
We see that, as expected, both expressions give rise to the 
same evolutions at early times (large redshifts) where they also agree
with the standard $\Lambda$CDM evolution. However, at late times the 
quasi-static approximation fails to correctly describe the evolution 
of perturbations.

Notice that the model example satisfies all the viability conditions 
described in the previous section except for the local gravity tests. 
As we will
show in the next section, it is precisely this last condition
$\vert f_{R}\vert\ll 1$ what will ensure the validity of the  quasi-static 
approximation.
 
\subsection{Recovering the quasi-static limit}

We will now restrict ourselves to models satisfying all the 
viability conditions, including $\vert f_{R}\vert\ll 1$. 

In Appendix I we have reproduced 
all the $\alpha$'s  and 
the first four $\beta$'s coefficients for each 
$\delta$ term in \eqref{delta_equation_separated}.  
These are the 
dominant ones for sub-Hubble modes (i.e. $\epsilon\ll1$) once 
the condition $\vert f_{R}\vert\ll 1$ has been imposed.
Thus, keeping only  $\sum_{j=1}^{4}\beta^{(j)}_{i=0,...,4,f}$ 
and $\alpha^{(1)}_{i=0,1,2,EH}$ as the relevant 
contributions for the general coefficients,  the full differential 
equation \eqref{delta_equation_separated} 
can be simplified as
\begin{eqnarray}
c_{4}\delta^{iv}+c_{3}\delta^{'''}
+c_{2}\delta^{''}+c_{1}\delta^{'}+c_{0}\delta\,=\,0
\end{eqnarray}
where coefficients $c's$ are written in Appendix II.

We see that indeed in the sub-Hubble limit the $c_4$ and
$c_3$ coeficients are negligible and the equation can be reduced to 
a second order expression.

As a consistency check, we find that both, 
in a matter dominated universe 
and in  $\Lambda \text{CDM}$ all $\beta$ 
coefficients vanish identically  since $f_1, f_2 \equiv 0$. 
For these cases, equation \eqref{delta_equation_separated} becomes 
equation \eqref{delta_0_SubHubble} as expected. For instance, 
in the pure matter dominated case, coefficients $\kappa$'s are constant 
and they take the following values $\kappa_1\,=\,-1/2$,  
$\kappa_2\,=\,1/2$, $\kappa_3\,=\,-3/4$ and $\kappa_4\,=\,3/2$.

Another important feature from our results is that, in general, 
without imposing $\vert f_{R}\vert\ll 1$,  
the quotient  
$(\alpha_{1, \text{EH}}+\beta_{1, f})$/
$(\alpha_{2, \text{EH}}+\beta_{2, f})$ is 
not always equal to $\mathcal{H}$. In fact 
only the quotients  
$\alpha^{(1)}_{1, \text{EH}}$/$\alpha^{(1)}_{2, \text{EH}}$ 
and $\beta^{(1)}_{1, f}$/$\beta^{(1)}_{2,f}$ are identically 
equal to $\mathcal{H}$  what is in agreement with the $\delta'$ 
coefficient in \eqref{delta_0_SubHubble}. 
However for our approximated 
expressions it is true that
$c_1/c_2\,\equiv\,\mathcal{H}$

From expressions in Appendix II, the 
second order equation for $\delta$ becomes
\begin{widetext}
\begin{eqnarray}
\delta''+\mathcal{H}\delta'-\frac{4}{3}\frac{\big[\frac{6 f_{RR}k^2}{a^2}+\frac{9}{4}(1-\sqrt{1-\frac{8}{9}\frac{2\kappa_1-\kappa_2}{-2+\kappa_2}})\big]\big[\frac{6 f_{RR}k^2}{a^2}+\frac{9}{4}(1+\sqrt{1-\frac{8}{9}\frac{2\kappa_1-\kappa_2}{-2+\kappa_2}})\big]}{
\big[\frac{6f_{RR}k^2}{a^2}+\frac{5}{2}(1-\sqrt{1-\frac{24}{25}\frac{-1+\kappa_1}{-2+\kappa_2}})\big]\big
[\frac{6 f_{RR}k^2}{a^2}+\frac{5}{2}(1+\sqrt{1-\frac{24}{25}\frac{-1+\kappa_1}{-2+\kappa_2}})\big]}(1-\kappa_1)\mathcal{H}^{2}\delta\,=\,0
\label{Qstatic_ours}
\end{eqnarray}
which can also be written as:
\begin{eqnarray}
\delta''+\mathcal{H}\delta'-\frac{4}{3}\frac{\big(\frac{6f_{RR}k^2}{a^2}+\frac{9}{4}\big)^{2}-\frac{81}{16}+\frac{9}{2}\frac{2\kappa_1-\kappa_2}{-2+\kappa_2}}{\big(\frac{6f_{RR}k^2}{a^2}+\frac{5}{2}\big)^{2}-\frac{25}{4}+6\frac{-1+\kappa_1}{-2+\kappa_2}}(1-\kappa_1)\mathcal{H}^{2}\delta\,=\,0
%{\big(\frac{6 %f_{RR}k^2}{a^2}+\frac{5}{2}\big)^{2}-\frac{25}{4}(1-\frac{25}{24}\frac{-1+\kappa_1}{-2+\kappa_2}})}\t%ilde{\rho}\delta\,=\,0
\label{Qstatic_ours_2}
\end{eqnarray}
\end{widetext}
Note that the quasi-static 
expression \eqref{Qstatic_strong_equation} 
is only recovered in the matter era (i.e. for $\mathcal{H}=2/\eta$) or for
a pure  $\Lambda\text{CDM}$ evolution  for the background dynamics. 
Nevertheless in the considered limit $\mid f_{R}\mid\ll1$ 
it can be proven using the background equations of motion that
\begin{eqnarray}
1+\kappa_1-\kappa_2 \approx 0
\end{eqnarray}
and therefore 
$2\kappa_1-\kappa_2 \approx -2+\kappa_2 \approx -1 
+ \kappa_1$ what allows to simplify 
expression \eqref{Qstatic_ours_2} to 
approximately become \eqref{Qstatic_strong_equation}.
This is nothing but the fact that for viable models 
the background evolution resembles that of $\Lambda\text{CDM}$
\cite{Silvestri_09_2007}.

In other words, although for general
$f(R)$ functions the quasi-static approximation is not justified, for 
those viable  functions describing the present phase of 
accelerated expansion and  satisfying local gravity tests, 
it gives a correct description for the evolution of perturbations.

%%%%%%%%%%%%%%%%%%%%%%%%%%%%%%%%%%%%%%%%%%%%%%%%%%%%%%%%%%%%%%%%%%%%%%%%%%%%%%%%
%%%%%%%%%%%%%%%%%%%%%%%%%%%%%%%%%%%%%%%%%%%%%%%%%%%%%%%%%%%%%%%%%%%%%%%%%%%%%%%%
\section{Some proposed models}
In order to check the results obtained  in the 
previous section, we propose two particular $f(R)$ theories which 
allow us to determine - at least numerically - all 
the quantities involved in the calculations and therefore to obtain 
solutions for \eqref{delta_equation_separated}. 
As commented before, for viable models the background 
evolution resemble that of $\Lambda \text{CDM}$ 
at low redshifts  and that of a matter dominated universe 
at high redshifts, i.e. 
the quantity $(R+f(R))/R$ tends to one 
in the high curvature regime. Nevertheless the $f(R)$ contribution 
gives the dominant contribution to the gravitational action for 
small curvatures and therefore it may explain the cosmological 
acceleration.
For the sake of concreteness we will fix the model 
parameters imposing a deceleration parameter today 
$q_{0} \approx -0.6$.

Thus, our first model (A) will be: $f(R) \,=\, c_{1} R^p$.
According to the results presented 
in \cite{Amendola} and \cite{Sawicki} viable models 
of this type  include both matter dominated and 
late-time accelerated universe provided the parameters 
satisfy $c_{1} < 0$ and $0 < p < 1$.  
We have chosen $c_1 \,=\,-4.3$ and $p\,=\,0.01$ 
in $H_0^2$ units. 
This choice does 
verify all the viability conditions, including 
$\vert f_{R} \vert \ll 1$ today.
For the second model (B): $f(R)\,=\,\frac{1}{c_1 R^{e_1}+c_2}$, 
we have chosen $c_1\,=\,2.5\cdot 10^{-4}$\,,\,$e_1\,=\,0.3$ 
and $c_2\,=\,-0.22$ also in the same units.

For each model, we compare our result (\ref{Qstatic_ours}) with 
the standard $\Lambda\text{CDM}$ and the quasi-static approximation
 \eqref{Qstatic_strong_equation} (see Figs. 2 and 3).
In both cases, the initial conditions are given at redshift 
$z = 1000$ where $\delta$ is assumed to behave as in a 
matter dominated universe, i.e. $\delta_{k}(\eta)\propto a(\eta)$ 
with no k-dependence.  We see that for both models, 
the quasi-static approximation gives a correct
description for the evolution which clearly deviates from the $\Lambda$CDM
case.  

In figure \eqref{Figure_Models_delta_vs_k} 
the density contrast evaluated today was plotted  as a 
function of $k$ for 
both models. The growing dependence of $\delta$ with respect to 
$k$ is verified. 
This modified $k$-dependence with respect to the 
standard matter dominated universe could give rise to 
observable consequences in the matter power spectrum, as shown in
\cite{Starobinsky2}, and could be used to constrain or even discard $f(R)$
theories for cosmic acceleration.

\begin{figure}[h]
\begin{center}
%\epsfxsize=10cm   %width of figure - will enlarge/reduce the figures
%\epsfbox{fig3.eps}
%\figurebox{2cm}{3cm}{} %to have a box alone
\resizebox{8.8cm}{6.4cm} 
%\resizebox{8.5cm}{!}
{\includegraphics{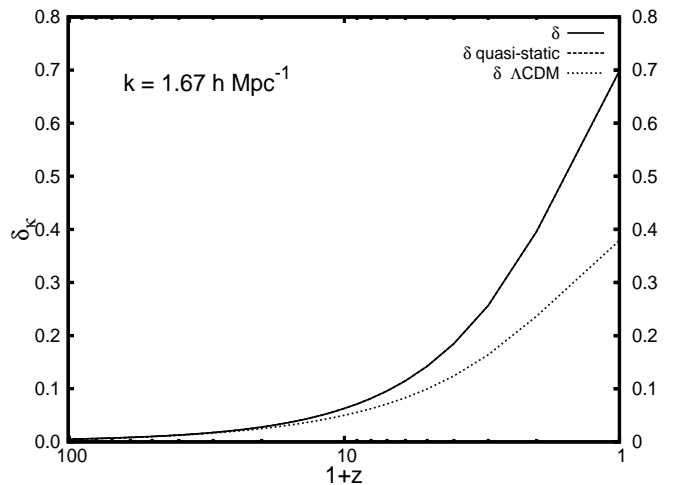}}
\caption {\footnotesize $\delta_{k}$ with $k = 1.67 
\text{h}\text{Mpc}^{-1}$ for $f(R)$ model $\bf{A}$ evolving 
according to 
\eqref{Qstatic_ours}, 
$\Lambda\text{CDM}$  and quasi-static approximation 
given by equation \eqref{Qstatic_strong_equation} 
in the redshift range from 1000 to 0. The quasi-static 
evolution is indistinguishable from that coming from 
(\ref{Qstatic_ours}), but diverges 
from $\Lambda\text{CDM}$ behaviour as $z$ decreases.}
\label{Figure_Model_2_k5000}
\end{center}
\end{figure}

\begin{figure}[h]
\begin{center}
%\epsfxsize=10cm   %width of figure - will enlarge/reduce the figures
%\epsfbox{fig3.eps}
%\figurebox{2cm}{3cm}{} %to have a box alone
\resizebox{8.8cm}{6.4cm} 
%\resizebox{8.5cm}{!}
{\includegraphics{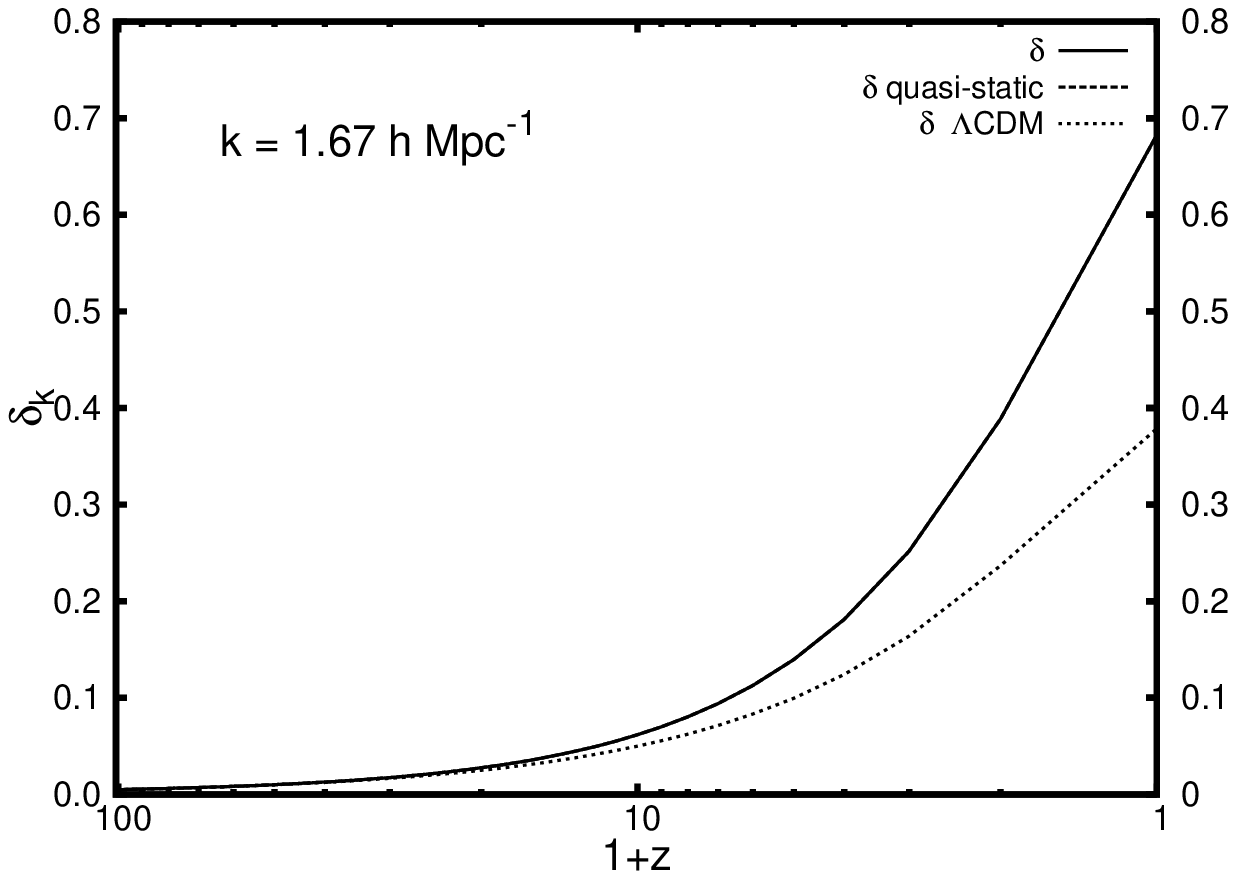}}
\caption {\footnotesize $\delta_{k}$ with $k = 1.67 
\text{h}\text{Mpc}^{-1}$ for $f(R)$ model $\bf{B}$ evolving according to 
\eqref{Qstatic_ours}, 
$\Lambda\text{CDM}$  and quasi-static evolution 
given by equation \eqref{Qstatic_strong_equation}
 in the redshift range from 1000 to 0. The quasi-static 
evolution is indistinguishable from that coming from 
(\ref{Qstatic_ours}), but diverges 
from $\Lambda\text{CDM}$ behaviour as $z$ decreases.}
\label{Figure_Model_3_k5000}
\end{center}
\end{figure}
\begin{figure}[h]
\begin{center}
%\epsfxsize=10cm   %width of figure - will enlarge/reduce the figures
%\epsfbox{fig3.eps}
%\figurebox{2cm}{3cm}{} %to have a box alone
\resizebox{8.8cm}{6.4cm} 
%\resizebox{8.5cm}{!}
{\includegraphics{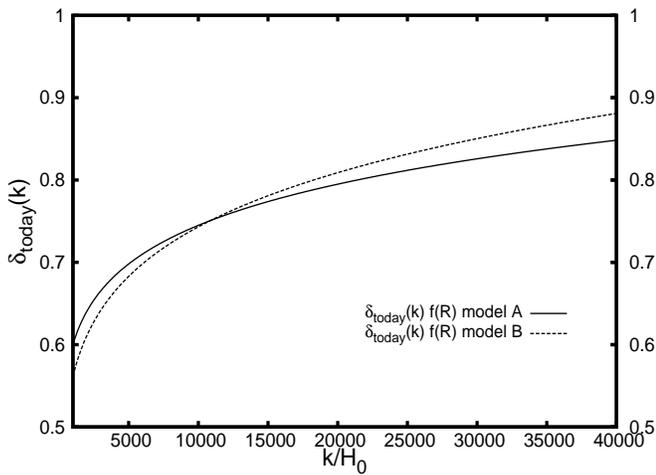}}
\caption {\footnotesize Scale dependence of $\delta_{k}$ evaluated 
today $(z=0)$ for $k/H_{0}$ in the  range from 1000 to 40000.}
\label{Figure_Models_delta_vs_k}
\end{center}
\end{figure}

\section{Conclusions}
In this work we have studied the evolution of matter density perturbations
in $f(R)$ theories of gravity. We have presented a completely general 
procedure to obtain the exact fourth-order differential equation
for the evolution of perturbations. We have shown that for sub-Hubble 
modes, the expression reduces to a second order equation. We have compared
this result with that obtained within quasi-static approximation 
used in the literature and found that for arbitrary $f(R)$ functions, such
an approximation is not justified. 

However, if we limit ourselves to theories
for which $\vert f_R\vert \ll 1$ today, then the perturbative calculation for sub-Hubble
modes requires to take into account, not only the first terms, 
but also higher-order terms in $\epsilon={\cal H}/k$. In that case, 
the resummation of such terms modifies the equation
which can be seen to be equivalent to the quasi-static case but only if
the universe expands as in  
a matter dominated phase or in a $\Lambda$CDM model. Finally, the fact 
that for models with $\vert f_R\vert \ll 1$ the background behaves today
precisely  as that of  $\Lambda$CDM makes the quasi-static approximation
 correct in those cases.
\vspace{.2cm}

{\bf Acknowledgements:} We would like to thank A. Starobinsky 
for useful comments and
J.\,A.\,R.\,Cembranos and J.\,Beltr\'an for their continuous 
encouragement. This work has been partially supported by the
 DGICYT (Spain) under projects FPA 2004-02602, FPA
2005-02327, CAM/UCM 910309 and by UCM-Santander PR34/07-15875.

\newpage
\begin{widetext}

%%%%%%%%%
\section{Appendices}
\subsection{Appendix $\bf{I}$: $\alpha's$ and $\beta's$ coefficients}
Coefficients for $\delta^{iv}$ term:
\begin{eqnarray}
\beta^{(1)}_{4,f}\,&\simeq&\,8f_{R}^{4}(1+f_{R})^{6}f_{1}^{4}\epsilon^{2}\nonumber\\
\beta^{(2)}_{4,f}\,&\simeq&\,72f_{R}^{3}f_{1}^{3}\epsilon^{4}(-2+\kappa_2)\nonumber\\
\beta^{(3)}_{4,f}\,&\simeq&\,216f_{R}^{2}f_{1}^{2}\epsilon^{6}(-2+\kappa_{2})^{2}\nonumber\\
\beta^{(4)}_{4,f}\,&\simeq&\,216f_{R}f_{1}\epsilon^{8}(-2+\kappa_{2})^{3}
\label{delta_4_coefficients}
\end{eqnarray}

Coefficients for $\delta'''$ term:
\begin{eqnarray}
\beta^{(1)}_{3,f}\,&\simeq&\,8f_{R}^{4}(1+f_{R})^{5}f_{1}^{4}
\mathcal{H}\epsilon^{2}[3+f_{R}(3+f_{1})]\nonumber\\
\beta^{(2)}_{3,f}\,&\simeq&\,6f_{R}^{3}f_{1}^{2}\mathcal{H}\epsilon^{4}
\{8f_{2}(-2+\kappa_{2})+4f_{1}[12\kappa_{1}+9\kappa_{2}-2(9+\kappa_{3})]\}\nonumber\\
\beta^{(3)}_{3,f}\,&\simeq&\,-72f_{R}^{2}f_{1}\mathcal{H}\epsilon^{6}
(-2+\kappa_{2})[-4f_{2}(-2+\kappa_{2})+f_{1}(19-23\kappa_{1}-10\kappa_{2}
+4\kappa_{3})]\nonumber\\
\beta^{(4)}_{3,f}\,&\simeq&\,-216f_{R}\mathcal{H}\epsilon^{8}(-2+\kappa_{2})^{2}[-2f_{2}(-2+\kappa_{2})+f_{1}(7-11\kappa_{1}-4\kappa_{2}+2\kappa_{3})]
\label{delta_3_coefficients}
\end{eqnarray}

Coefficients for $\delta''$ term:
\begin{eqnarray}
\alpha^{(1)}_{2,\text{EH}}\,&=&\,432(1+f_{R})^{10} \mathcal{H}^{2} \epsilon^{8} (-1+\kappa_1)(-2+\kappa_2)^{3}\nonumber\\
\alpha^{(2)}_{2,\text{EH}}\,&=&\,1296(1+f_{R})^{10} \mathcal{H}^{2} \epsilon^{10} (-1+\kappa_1)^{2}(-2+\kappa_2)^{3}\nonumber\\
\alpha^{(3)}_{2,\text{EH}}\,&=&\,3888(1+f_{R})^{10} \mathcal{H}^{2}\epsilon^{12}(-1+\kappa_{1})^{2}(-2+\kappa_2)^{3}\nonumber\\
\beta^{(1)}_{2,f}\,&\simeq&\,8f_{R}^{4}(1+f_{R})^{6}f_{1}^{4}\mathcal{H}^{2}\nonumber\\
\beta^{(2)}_{2,f}\,&\simeq&\,88f_{R}^{3}f_{1}^{3}\mathcal{H}^{2}\epsilon^{2}(-2+\kappa_2)\nonumber\\
\beta^{(3)}_{2,f}\,&\simeq&\,24f_{R}^{2}f_{1}^{2}\mathcal{H}^{2}\epsilon^{4}(-2+\kappa_2)(-28+2\kappa_{1}+13\kappa_{2})\nonumber\\
\beta^{(4)}_{2,f}\,&\simeq&\,72f_{R}f_{1}\mathcal{H}^{2}\epsilon^{6}(-2+\kappa_2)^{2}(-14+4\kappa_{1}+5\kappa_{2})
\label{delta_2_coefficients}
\end{eqnarray}

Coefficients for $\delta'$ term:
\begin{eqnarray}
\alpha^{(1)}_{1,\text{EH}}\,&=&\,432(1+f_{R})^{10} \mathcal{H}^{3} \epsilon^{8} (-1+\kappa_1)(-2+\kappa_2)^{3}\nonumber\\
\alpha^{(2)}_{1,\text{EH}}\,&=&\,2592(1+f_{R})^{10} \mathcal{H}^{3} \epsilon^{10}(-1+\kappa_1)^{2}(-2+\kappa_{2})^{3}\nonumber\\
\alpha^{(3)}_{1,\text{EH}}\,&=&\,-7776(1+f_{R})^{10} \mathcal{H}^{3}\epsilon^{12}(-1+\kappa_{1})^{3}(-2+\kappa_{2})^{3}\nonumber\\
\beta^{(1)}_{1,f}\,&\simeq&\,8f_{R}^{4}(1+f_{R})^{6}f_{1}^{4}\mathcal{H}^{3}\nonumber\\
\beta^{(2)}_{1,f}\,&\simeq&\,88f_{R}^{3}f_{1}^{3}\mathcal{H}^{3}\epsilon^{2}(-2+\kappa_2)\nonumber\\
\beta^{(3)}_{1,f}\,&\simeq&\,24f_{R}^{2}f_{1}^{2}\mathcal{H}^{3}\epsilon^{4}(-2+\kappa_2)(-28+2\kappa_{1}+13\kappa_{2})\nonumber\\
\beta^{(4)}_{1,f}\,&\simeq&\,72f_{R}f_{1}\mathcal{H}^{3}\epsilon^{6}(-2+\kappa_2)^{2}(-14+4\kappa_{1}+5\kappa_{2})
\label{delta_1_coefficients}
\end{eqnarray}

Coefficients for $\delta$ term:
\begin{eqnarray}
\alpha^{(1)}_{0,\text{EH}}\,&=&\,432(1+f_{R})^{10} \mathcal{H}^{4} \epsilon^{8} (-1+\kappa_1)(2\kappa_1-\kappa_2)(-2+\kappa_2)^{3}\nonumber\\
\alpha^{(2)}_{0,\text{EH}}\,&=&\,1296(1+f_{R})^{10}\mathcal{H}^{4} \epsilon^{10}(-1+\kappa_1)^{2}(-1+4\kappa_1-\kappa_2)(-2+\kappa_{2})^{3}\nonumber\\
\alpha^{(3)}_{0,\text{EH}}\,&=&\,3888(1+f_{R})^{10} \mathcal{H}^{4}\epsilon^{12}(-1+\kappa_{1})^{2}(2\kappa_{1}^{2}-\kappa_2)(-2+\kappa_{2})^{3}\nonumber\\
\beta^{(1)}_{0,f}\,&\simeq&\,-\frac{16}{3}f_{R}^{4}(1+f_{R})^{5}f_{1}^{4}\mathcal{H}^{4}
[2+f_{R}(2+2f_{1}-f_{2}-2\kappa_1)-2\kappa_1]\nonumber\\
\beta^{(2)}_{0,f}\,&\simeq&\,112f_{R}^{3}f_{1}^{3}\mathcal{H}^{4}\epsilon^{2}(-1+\kappa_{1})(-2+\kappa_{2})\nonumber\\
\beta^{(3)}_{0,f}\,&\simeq&\,48f_{R}^{2}f_{1}^{2}\mathcal{H}^{4}\epsilon^{4}(-1+\kappa_{1})(-2+\kappa_{2})(-16+2\kappa_{1}+7\kappa_{2})\nonumber\\
\beta^{(4)}_{0,f}\,&\simeq&\,144f_{R}f_{1}\mathcal{H}^{4}\epsilon^{6}(-1+\kappa_{1})(-2+\kappa_{2})^{2}(-6+4\kappa_{1}+\kappa_{2})
\label{delta_0_coefficients}
\end{eqnarray}
%%%%%%%%%%
\subsection{Appendix $\bf{II}$: $c's$ coefficients}
\begin{eqnarray}
c_{4}\,&=&\,-f_{R}f_{1}[-f_{R}f_1k^{2}-3\mathcal{H}^2(-2+\kappa_2)]^{3}\nonumber\\
c_{3}\,&=&\,-3f_{R}\mathcal{H}[-f_{R}f_1k^2-3\mathcal{H}^2(-2+\kappa_2)]\{
f_{R}^{2}f_{1}^{3}k^4+6f_2\mathcal{H}^4(-2+\kappa_2)^2+f_1\mathcal{H}^2(-2+\kappa_2)[2f_{R}f_2k^2+\nonumber\\
&&3\mathcal{H}^2(-7+11\kappa_1+4\kappa_2-2\kappa_3)]+2f_{R}f_1\mathcal{H}^2k^2(-6+6\kappa_1+3\kappa_2-\kappa_3)\}\nonumber\\
c_{2}\,&=&\,[-f_{R}f_1k^2-3\mathcal{H}^2(-2+\kappa_2)]^{2}[f_{R}^{2}f_{1}^{2}k^4+5f_{R}f_1\mathcal{H}^2k^2(-2+\kappa_2)+6\mathcal{H}^4(-1+\kappa_1)(-2+\kappa_2)]\nonumber\\
c_{1}\,&=&\,\mathcal{H} c_{2}\nonumber\\
c_{0}\,&=&\,\frac{2}{3}\mathcal{H}^2(-1+\kappa_1)
[-f_{R}f_1k^2-3\mathcal{H}^2(-2+\kappa_2)]^{2}
[2f_{R}^{2}f_{1}^{2}k^4+9f_{R}f_1
\mathcal{H}^2k^2(-2+\kappa_2)+9\mathcal{H}^4
(2\kappa_1-\kappa_2)(-2+\kappa_2)]\nonumber \\
\label{coefficients_cs}
\end{eqnarray}
\end{widetext}

\end{document}